\documentclass[%
 reprint,
 amsmath,amssymb,
 aps,
]{revtex4-2}

\usepackage{graphicx}
\usepackage{dcolumn}
\usepackage{bm}
\usepackage{upgreek}

\begin{document}

\preprint{APS/123-QED}

\title{Gain-Bandwidth-Product-Induced Technical Bound in Time Transfer System \\ without Inline Amplifiers}

\author{Yufei Zhang}
\author{Ziyang Chen}
\email{chenziyang@pku.edu.cn}
\author{Hong Guo}
\email{hongguo@pku.edu.cn}
\affiliation{State Key Laboratory of Advanced Optical Communication Systems and Networks, School of Electronics, and Center for Quantum Information Technology, Peking University, Beijing 100871, China}

\date{\today}

\begin{abstract}
Time transfer plays a dispensable role in many fields including navigation and positioning, geodesy, and fundamental tests. However, in certain scenarios where effective relay node deployment is often not feasible, such as harsh environments with extremely poor infrastructure and emergency conditions, effective time transfer without inline amplifiers becomes crucial. In this situation, the maximum transmission distance is limited by the receiver's measurement capability, particularly its ability to amplify the signal. Here we propose a theoretical model, giving a technical lower bound of the detected signal stability at different transmission distances, induced by limited gain-bandwidth products. The results under common gain-bandwidth products show that while for shorter transmission distances, stability is mainly limited by the background noise of the time interval counter, for longer distances reaching the scale of 300 kilometers, the technical lower bound is below the level of 10 nanoseconds without any inline amplification devices. Therefore, the given technical bound offers guidance on managing the balance between distance and stability, together with the optimization of the receiver in long-distance time transfer without inline amplification.
\end{abstract}

\maketitle


\section{\label{sec:level1}Introduction}
Time and frequency are currently the most precisely measured physical quantities. In many fields of scientific research and practical applications, high-precision time standards are essential. To date, atomic clocks have evolved into various types, including cesium clocks, rubidium clocks, hydrogen clocks, etc. The typical instability of microwave clocks has reached the order of $10^{-14}$\cite{RN720}, while optical clocks have achieved instability in the order of $10^{-18}$\cite{RN106}. However, for high-precision atomic clocks especially the emerging optical clocks, limited by the cost and complexity, they can only be set up in a few locations and must provide time standards to a wide range of users through time dissemination services. This is indispensable in a large variety of fields, to name a few, navigation and positioning\cite{RN702, RN8, RN712}, geodesy\cite{RN713, RN714}, search for dark matters\cite{RN705, RN706} and fundamental tests\cite{RN7, RN703, RN706}. Moreover, high-precision optical clock comparisons\cite{RN733, RN737, RN738, RN112, RN704} also require long-distance transmission of clock signals, making high-precision time transfer crucial. The phase information of signals in optical fibers remains relatively stable, providing a more reliable environment for long-distance transmission. Additionally, optical fiber networks are vital infrastructure across wide terrestrial areas, making long-distance time and frequency transfer via optical fibers very promising. Significant progress have been made in this field in the past years\cite{RN707, chen2024, RN737, RN735, RN701, RN676, RN709, RN742, RN42, RN698, RN43, RN694, RN736, RN739, RN737}.

In optical fiber time transfer, a two-way comparison method is used, where two pulse trains are sent in both directions at the source and the remote ends. By measuring the time interval between the received and transmitted signals, clock difference information can be obtained and compensated for, ultimately achieving high-precision time transfer. The accuracy of time transfer primarily depends on the detection capability of the transmitted signals at both the source and remote ends and the precision of the time interval measurements. Traditionally, at the receiving end, an optical detector converts the optical signal to an electrical signal, which is then input into a time interval counter (TIC) to measure the time interval. Recently, schemes based on linear optical sampling\cite{RN21, RN12} have further improved measurement accuracy\cite{chen2024, RN673, RN744, RN734, RN607, RN730, RN3, RN729, RN710, RN741, RN740, RN14}. In optical fiber links, cascading relay nodes are typically added to the link to compensate for effects like signal attenuation and dispersion\cite{RN707, RN676}, ensuring the quality of the received signal. There are two main types of relay nodes: optical relays and electrical repeaters. Optical relays achieve long-distance signal transmission by receiving the optical signal, replenishing optical power, and retransmitting the optical signal successively\cite{RN735, RN707}. Electrical relays perform optical-electrical conversion, electrical amplification, and electro-optical conversion in turn to enhance signal power equivalently and realize long-distance time transfer\cite{RN708, RN709}.

However, in certain special but critical application scenarios, effective relay node deployment is often not feasible. For examples, in harsh environments with extremely poor infrastructure, setting up relay nodes is challenging; in deserts, plateaus, polar regions, underwater, and other areas lacking signal regeneration technology, optical cables can only support point-to-point transmission; in emergent situations, there is not enough time to set up relay stations. These scenarios urgently require time transfer technique without inline amplification. On the other hand, the longer the transmission distance, the greater the attenuation, necessitating higher amplification at the receiving end when using traditional detection methods. The gain at the receiving end ultimately limits whether the transmitted signal can be effectively received, while the bandwidth determines whether the waveform can be faithfully received without distortion. However, there is a trade-off between these two important quantities, where the limited gain-bandwidth product (GBP)\cite{RN717, RN716, RN719} will significantly sacrifice the detection bandwidth under large amplification, and thus degrade the quality of the received signal, reducing time transfer accuracy. Therefore, in those time transfer scenarios without inline amplifiers, it's necessary to study the technical limits of time transfer over certain distances to provide important references for practical relay-free single-span applications.

Here we present a theoretical model in absence of inline optical amplifiers for time transfer, in consideration of the limited GBP, which is often the case in real experiments. We conduct analyses on the impacts of detector noise and the limited voltage resolution of the TIC on the stability of the detected signal under different bandwidths. By establishing the relationship between transfer distance and maximum effective detection bandwidth under certain GBPs, we provide the technical lower bounds on the instability of the detected signal at different transfer distances. The simulation results show that for shorter transfer distances, the stability of the detected signal is primarily limited by the time resolution of TIC; for longer transfer distances, it is mainly constrained by the limited GBP. As the transfer distance increases, the lower bound at 300-km scale can only reach the level of 10 nanoseconds, implying that the positioning accuracy based on time transfer can only achieve the order of one meter. Therefore, new detection methods with higher GBP inevitably need to be adopted, in order to reach enough precision for practical applications.

This paper is organized as follows: Section~\ref{Sec2} presents the model of time transfer system without inline amplifiers; Section~\ref{Sec3} provides the simulation and experimental results, along with analyses and discussions; and Section~\ref{Sec4} concludes the paper.

\section{Model of Time Transfer System without Inline Amplifiers\label{Sec2}}
\begin{figure}
\includegraphics[scale = 0.88]{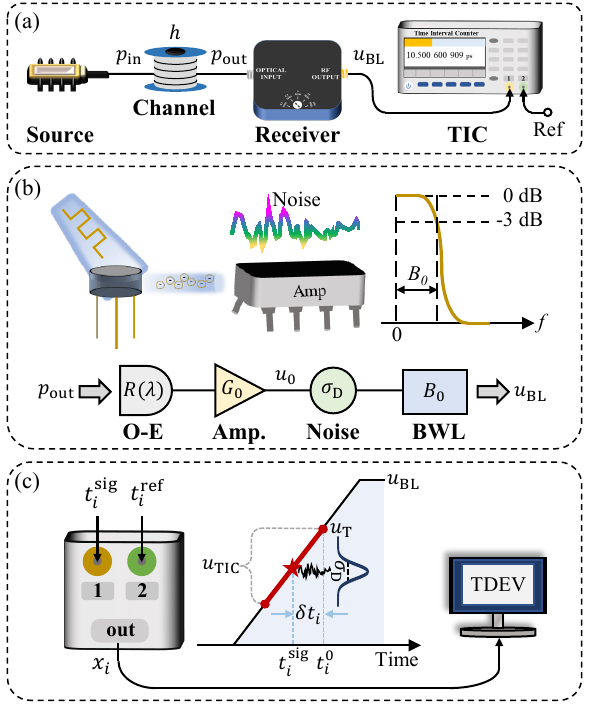}
\caption{\label{Model} Model of time transfer system without inline amplifiers. (a) Overview of the system, where the optical source, channel, receiver and TIC modules are included. (b) Schematic of the receiver. The income optical pulse is converted into currents with noise added. Under a certain value of GBP, there is a trade-off between amplification and effective bandwidth. (c) Process of time interval measurement, where the signal and its reference are connected to the two inputs of TIC to measure the time interval, and generate the TDEV via data processing. The limited voltage resolution and the additive noise, together with their impacts on the time domain are also pictured here.}
\end{figure}

The model of time transfer without inline amplifiers is split into five parts: pulse transmission, GBP-limited pulse measurement, trigger count of the signal, measurement points determination, and instability evaluation. These are analyzed detailedly in the following context with the schematic shown in Fig.~\ref{Model}. 

\subsection{Pulse Transmission}
We delve into the propagation of the optical pulse sequence $p_\text{in}(t)$ in the channel. The clock information is carried by the optical pulse train with a period $T$. We denote the transfer function of the channel as $h(t)$, and it can be written as
\begin{equation}
    h(t) = 10^{-\alpha L/10}\times h^\text{dis}(t).
    \label{Eq1}
\end{equation}
where we currently take the attenuation and dispersion of the channel into account. The first term on the right side of Eq.~(\ref{Eq1}) represents the channel attenuation, where $\alpha$ [dB/km] is the channel attenuation coefficient and $L$ stands for the channel length. The second term represents the channel dispersion. Therefore, the signal after passing through the channel is given by
\begin{equation}
    p_\text{out}(t)=[p_\text{in}(t)\ast h^\text{dis}(t)]\times10^{-\alpha L/10}.
    \label{Eq2}
\end{equation}
\subsection{GBP-Limited Pulse Measurement}
The optical signal transmitted through the channel enters the receiver module, with the process of photodetection and amplification. The output of this process is an electrical signal, satisfying
\begin{equation}
    u_0(t)=p_\text{out}(t)\times R(\lambda)\times Z_\text{TIA}\times G_0,
    \label{Eq3}
\end{equation}
where $p_\text{out}(t)$ represents the signal input to the receiver module after passing through the channel, $R(\lambda)$ [A/W] is the responsivity of the photodiode at wavelength $\lambda$, $Z_\text{TIA}$ is the output impedance of the transimpedance amplifier, and $G_0$ [dB] is the electrical gain provided by the receiver module. For time interval measurements, it is required that the output power of the electrical signal at the detection end meets the requirements of subsequent measurement and usage. Therefore, in the case of single-span transmission without relay amplifiers, the electrical amplification at the detection end needs to compensate for the attenuation introduced by the channel as much as possible, which is
\begin{equation}
    G_0 = \eta \times 10^{\frac{\alpha L}{10}}.
    \label{Eq4}
\end{equation}

The effective amplification efficiency $\eta > 0$ reflects the effective amplification capability of the receiver module. When $\eta > 1$, the power of the signal increases. Generally, $\eta = 1$ is used for calculation, indicating that the attenuation of signal intensity by the channel is compensated for by amplification.

However, at the same time, the receiver module has a finite gain bandwidth product $\beta$ [Hz], which creates a trade-off relationship between the maximum gain and effective bandwidth at the detection end, thereby limiting the maximum transmission distance without inline amplification. The gain bandwidth product is defined as\cite{RN715}
\begin{equation}
    \beta=G\times B,
    \label{Eq5}
\end{equation}
where $B$ represents the effective bandwidth of the receiver module. Therefore, combined with Eq.~(\ref{Eq4}), to achieve sufficient electrical signal power, the maximum effective bandwidth that the receiver module can provide under the amplification factor $G_0$ is given by
\begin{equation}
    B_0 = \frac{\beta}{\eta} \times 10^{-\frac{\alpha L}{10}}.
    \label{Eq6}
\end{equation}

Without loss of generality, we use a window function $W(f)$ with a width of $B_0$ to represent the effect of limited detection bandwidth, i.e.,
\begin{equation}
    W(f) = \begin{cases} 1, & |f| < B_0 \\ 0. & |f| > B_0 \end{cases}
    \label{Eq7}
\end{equation}
Therefore, the band-limited signal is
\begin{equation}
    u_\text{BL}(t) = 2B_0u_0(t) \ast \text{sinc}(2B_0 t),
    \label{Eq8}
\end{equation}
where the sinc function is defined as $\text{sinc}\theta = \sin(\pi \theta)/\pi\theta$. Substituting Eqs.~([\ref{Eq2}-\ref{Eq4}, \ref{Eq6}]) into Eq.~(\ref{Eq8}), we obtain the band-limited electrical signal output by the receiver module under single-span channel without inline amplification as
\begin{eqnarray}
    u_\text{BL}(t) =&& 2\beta R(\lambda)Z_{\text{TIA}}\times10^{-\frac{\alpha L}{10}} \nonumber\\
    &&\times\left[P_{\text{in}}(t) \ast h^{\text{dis}}(t) \ast \text{sinc}\left(\frac{2\beta t}{\eta}\times10^{-\frac{\alpha L}{10}}\right)\right].
\label{Eq9} 
\end{eqnarray}
And the process described above is depicted in Fig.~\ref{Model}(b).

\subsection{Trigger Count of the Signal}
We input it into one channel of TIC the final output band-limited signal $u_{\text{BL}}(t)$ of the receiver module. In total, we measure a continuous sequence of $N$ periods of pulses to complete the measurement and evaluation of stability for different time scales of the received signal. The TIC requires setting a certain triggering voltage threshold $u_\text{T}$ to select measurement points, and sequentially completes the collection of the moments when the rising edges of the pulses arrive within each period. Taking the i-th period as an example, we denote the measurement time corresponding to when its rising edge voltage reaches $u_\text{T}$ as $\{t_i^{\text{sig}}\}~(i=1,2,\dots,N)$, which can be decomposed into a theoretical value and a noise term, i.e., 
\begin{equation}
    t_i^{\text{sig}} = t_i^0 + \delta t_i.
    \label{Eq10}
\end{equation}
The first term $t_i^0$ on the right-hand side of Eq.~(\ref{Eq10}) is the theoretical value of the time measured by the TIC when the rising edge arrives. It satisfies
\begin{equation}
    \begin{cases} 
    t_i^0 = u_\text{BL}^{-1}(u_\text{T}), \\ u_\text{T} = u_\text{BL}(t_i^0),
    \end{cases}
    \label{Eq11}
\end{equation}
where $u_\text{BL}^{-1}$ denotes the inverse function of the waveform. Meanwhile, since $u_\text{BL}(t)$ is a periodic function, we have
\begin{equation}
    t_{i+1}^0 - t_i^0 = T.
    \label{Eq12}
\end{equation}

The second term on the right-hand side of Eq.~(\ref{Eq10}), $\delta t_i$, is the error term introduced. It includes the PD noise $n_\text{D}(t)$, which is generally additive Gaussian white noise and the quantization noise $n_\text{Q}(t)$ introduced after taking the finite voltage resolution of the TIC into consideration\cite{RN727, RN726}. Next, we will analyze the statistical distributions satisfied by these noise and the specific influence they have.

\subsection{Measurement Points Determination}
The aforementioned two noises will both lead to certain voltage fluctuations. For different waveforms, the same voltage fluctuation will result in different measurement time errors. We assume that the voltage fluctuation caused by noise at each moment is denoted as $u_\text{noise}$, which follows a certain statistical distribution. Let $f(u)$ represent the probability density function of this statistical distribution. Then the probability that the voltage fluctuation at a certain moment is greater than $\delta u$ is given by 
\begin{equation}
P(u_\text{noise} \geq \delta u) = \int_{\delta u}^{\infty} f(u)du.
\label{Eq13}
\end{equation}

We consider the rising edge in detail. As shown in Fig.~\ref{Model}(c), for the moment in the rising edge satisfying $u_\text{BL} = u_\text{T} - \delta u$, as long as its voltage value satisfies $u_\text{BL} + u_\text{noise} \geq u_\text{T}$, reaching or exceeding the threshold voltage, TIC will complete the triggering of the signal rising edge and record the corresponding time $t_i^{\text{sig}}$. This requires that the voltage fluctuation value at that moment $u_\text{noise} \geq \delta u$, the probability of which has been given by Eq.~(\ref{Eq13}). Similar to Eq.~(\ref{Eq11}), we have
\begin{equation}
\begin{cases}
    t_i^{\text{sig}} = u_\text{BL}^{-1}(u_\text{T} - \delta u), \\ u_\text{T} - \delta u = u_\text{BL}(t_i^{\text{sig}}).
\end{cases}
\label{Eq14}
\end{equation}
Here we perform a series expansion near the voltage threshold, yielding
\begin{equation}
u_\text{BL}(t) = u_\text{BL}(t_i^0) + u^\prime_{\text{BL}}(t_i^0) \cdot (t - t_i^0) + \frac{1}{2} u^{\prime\prime}_{\text{BL}}(t_i^0) \cdot (t - t_i^0)^2 + \cdots,
\label{Eq15}
\end{equation}
where $u^\prime_{\text{BL}}(t_i^0)$ and $u^{\prime\prime}_{\text{BL}}(t_i^0)$ are respectively the first and second derivatives of $u_{\text{BL}}(t)$ at $t_i^0$. Since the voltage fluctuation is relatively small compared to the detection signal voltage, under a small-range linear approximation, substituting Eq.~(\ref{Eq14}) yields
\begin{equation}
t_i^{\text{sig}} = t_i^0-\frac{\delta u}{u^\prime_{BL}(t_i^0)}.
\label{Eq16}
\end{equation}
Compared with Eq.~(\ref{Eq10}), we obtain
\begin{equation}
\delta t_i = -\frac{\delta u}{u'_{\text{BL}}(t_i^0)}.
\label{Eq17}
\end{equation}
Meanwhile, combined with Eq.~(\ref{Eq13}), we derive that the probability of triggering at the moment with an error of $\delta t_i$ is
\begin{equation}
P_{\text{trigger}} = \int_{\delta t_i \cdot u^\prime_{\text{BL}}(t_i^0)}^{\infty} f(u)du.
\label{Eq18}
\end{equation}

Next, we specifically analyze the two types of errors introduced by the detector noise $n_\text{D}(t)$ and quantization noise $n_\text{Q}(t)$. Firstly, we consider the noise introduced by the detector $n_\text{D}(t)$. In general, it is additive white Gaussian noise, so the voltage fluctuation introduced follows a Gaussian distribution, i.e., in this case $u_{\text{Noise}} = n_\text{D}(t) \sim N(\mu_\text{D}, \sigma_\text{D}^2)$. Since we are considering the noise additive to the signal, the expectation value $\mu_\text{D} = 0$, and $\sigma_\text{D}^2$ reflects the noise variance. The specific noise intensity can be obtained by measuring the system noise background experimentally. Therefore, the probability density function of Gaussian noise can be written as
\begin{equation}
f_\text{D}(u) = \frac{1}{\sqrt{2\pi}\sigma_\text{D}}\exp\left(-\frac{u^2}{2\sigma_\text{D}^2}\right).
\label{Eq19}
\end{equation}

On the other hand, the TIC has a finite voltage resolution $u_{\text{TIC}}$, which generally leads to quantization noise that follows a uniform distribution, i.e., in this case $u_{\text{Noise}} = n_\text{Q}(t) \sim \text{U}\left(-u_\text{TIC}/2,u_\text{TIC}/2\right)$, and its probability density function is
\begin{equation}
f_\text{Q}(u) = \begin{cases} 
1/u_\text{TIC},&-u_\text{TIC}/2 < u < u_\text{TIC}/2\\
0.&\text{else}
\end{cases} 
\label{Eq20}
\end{equation}

In summary, Eqs.~([\ref{Eq17}, \ref{Eq18}]) have provided a complete description of the timing errors of signal acquisition caused by voltage noise under different waveforms. Eqs.~([\ref{Eq19}, \ref{Eq20}]) respectively give specific descriptions of the two types of noise we are concerned about, and other types of noise can also be described similarly by writing out the corresponding probability density functions.

\subsection{Instability Evaluation}
Previously, we mentioned that we input the received signal into one channel of TIC. In order to evaluate its instability, another channel of TIC is connected with a reference signal. The triggering moments of the reference signal are denoted as $\{t_i^\text{ref}\}(i=1,2,\cdots,N)$. Generally, due to the higher stability of the reference signal compared to the signal under test, we assume that the triggering moments of the reference signal in adjacent periods differ by the period $T$, i.e.,
\begin{equation}
t_{i+1}^\text{ref} - t_i^\text{ref} = T.
\label{Eq21}
\end{equation}

The TIC obtains a phase time difference sequence $\{ x_i \}(i=1,2,\ldots,N)$ by sequentially recording the differences between the detection signal and the reference signal triggering moments within each period, i.e.,
\begin{equation}
x_i = t_i^\text{sig} - t_i^\text{ref}.
\label{Eq22}
\end{equation}

The phase time difference sequence $\{x_i\}$ contains information about the instability of the transmitted signal. We generally use the time deviation (TDEV) as a quantitative measure, which is root of the time variance $\sigma_x^2(\tau)$. The smaller $\sigma_x^2(\tau)$ is, the higher stability the transmitted signal reaches. It is defined as
\begin{equation}
\sigma_x^2(\tau) = \frac{1}{6} \left\langle \left[ \frac{1}{n} \sum_{i=1}^{n} (x_{i+2n} - 2x_{i+n} + x_i) \right]^2 \right\rangle,
\label{Eq23}
\end{equation}
where $\tau = nT$ is the gate time, used to evaluate the instability of the received signal at different time scales, $n$ is a positive integer, and $\langle...\rangle$ indicates taking the expectation. Substituting Eqs.~([\ref{Eq10}, \ref{Eq12}, \ref{Eq21}-\ref{Eq22}]), we further obtain
\begin{equation}
    \sigma_x^2 (\tau) = \frac{1}{6} \left\langle \frac{1}{n^2} \left[ \sum_{i=1}^{n} (\delta t_{i+2n} - 2\delta t_{i+n} + \delta t_i) \right]^2 \right\rangle.
\label{Eq24}
\end{equation}

Since we have collected a total of $N$ consecutive periods, the expectation in Eq.~(\ref{Eq24}) can be explicitly expanded to obtain the expression
\begin{widetext}
\begin{equation}
    \sigma_x^2 (\tau) = \frac{1}{{6n^2(N-3n+1)}} \sum_{j=1}^{N-3n+1} \left[ \sum_{i=j}^{n+j-1} (\delta t_{i+2n} - 2\delta t_{i+n} + \delta t_i) \right]^2,
\label{Eq25}
\end{equation}
\end{widetext}
where $1 \leq n \leq \left\lfloor \frac{N}{3} \right\rfloor$. And Eqs.~([\ref{Eq17}, \ref{Eq18}]) connect $\delta t_i$ with the band-limited signal $u_\text{BL}$ and the noise probability density function $f(u)$.

\section{Results and Discussions\label{Sec3}}
After constructing the time transfer model without inline amplifiers as described above, in this section, we combine practical application scenarios and experimental conditions to conduct a detailed study of key parameters in the theoretical model through simulation methods. In experiments, we typically use one pulse per second signal (1 PPS) as the time reference and load it onto cw laser through modulation. Therefore, without loss of generality, in the following analyses, we set the signal period $T = 1$ s. We simulate and study the effects of the detection bandwidth $B_0$ on the detection waveform $u_{\text{BL}}$, the influence of detector noise $n_\text{D}$ and TIC voltage resolution $u_{\text{TIC}}$ on the stability of the transmitted signal, the technical lower bound of the instability of the transmitted signal at different transmission distances $L$ under some values of GBP $\beta$.

\subsection{Influences of detection bandwidth on the transmitted signal}
We start by analyzing the influences of the measurement bandwidth $B_0$ on the detection waveform $u_\text{BL}$ from the perspective of the signal receiver in the theoretical model. Equation~(\ref{Eq8}) provides the relationship between the measurement bandwidth $B_0$ and the band-limited signal $u_\text{BL}$. In our experiments, we use a photodetector based on an InGaSn PIN diode (Thorlabs PDB 450C), with typical adjustable detection bandwidths of $B_0=150$ MHz, 45 MHz, 4 MHz, 0.3 MHz, and 0.1 MHz. Therefore, we select these experimental parameters for simulation.

We use a second pulse with a pulse width of 10 $\upmu$s. To ensure sufficient accuracy, a high sampling rate of $10^{12}$ Hz is chosen for simulation, corresponding to a time interval of 1 ps. We analyze signals of length 100 $\upmu$s under the above conditions and simulate the signals obtained with different detection bandwidths.
\begin{figure}[b]
\includegraphics[scale = 0.85]{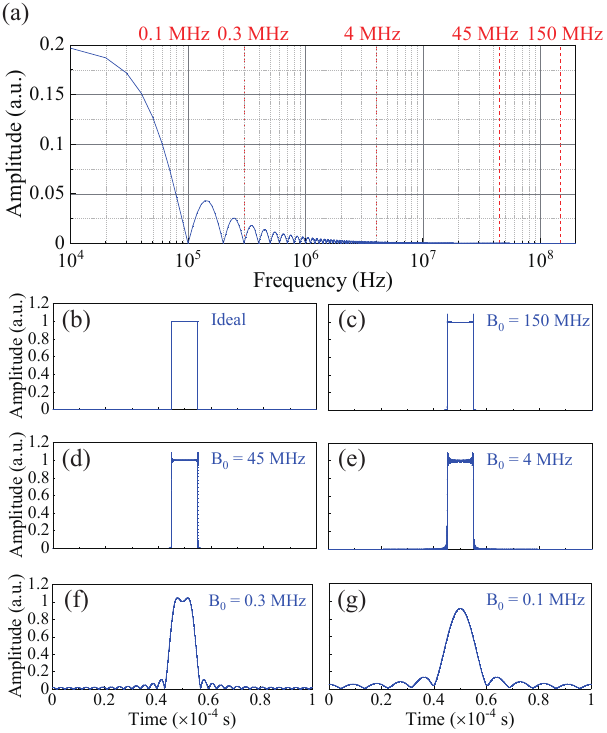}
\caption{Detected pulse signals under different detector's bandwidths. The frequency-domain of the signal is presented in subfigure (a) and the time-domain under different bandwidths are shown in subfigures from (b) to (g).}
\label{Different_Bandwidth} 
\end{figure}
Figure~\ref{Different_Bandwidth}(a) shows the single-sideband spectrum of a square wave signal, with a frequency resolution of $f_\text{s}/M=10^4$ Hz, where $f_\text{s}$ is the sampling rate and $M$ represents the total data length. Adding different bandwidth limitations results in the time-domain waveforms displayed in Fig.~\ref{Different_Bandwidth}(b)-(g). It can be observed that as the detection bandwidth decreases, the rising edge of the detected pulse signals gradually flattens, and the signal distorts progressively. The minimum bandwidth required to ensure undistorted detection signals is on the order of megahertz. As analyzed in Section~\ref{Sec2}, flatter rising edges correspond to lower signal stability in measurements. Therefore, for long-distance transmission and high amplification factors, a certain value of GBP inevitably leads to a reduction in detection bandwidth, thereby decreasing the measurement stability of the transmitted signal.

\subsection{Influences of different noises on the instability of the transmitted signal}

After analyzing the impact of effective bandwidth on the detection waveform, we specifically delve into the influence of noise $\delta t_i$ in Eq.~(\ref{Eq10}) on the final assessment of the instability of the transmitted signal under different detection bandwidths. In the experiment, the signal amplitude $u_\text{BL}$ obtained through detection is approximately 1 V. We normalize with respect to the signal amplitude for reference, thereby setting the intensity of the added detector Gaussian white noise $\sigma_\text{D}$ and the voltage resolution $u_\text{TIC}$. The detector's background noise in the receiver module is on the order of millivolts, differing by roughly three orders of magnitude; while the typical voltage resolution of the TIC (Agilent 53230A) used in the experiment is 20 $\text{mV}_\text{pk}$, differing by approximately two orders of magnitude from the signal amplitude. For signal acquisition, the experiment typically sets the trigger level at 50\% of the signal rising edge. Combining the above parameters, we simulate the process of continuous sampling 1000 times, i.e., $N=1000$, for different waveforms $u_\text{BL}$ obtained under different bandwidths $B_0$. We simulated under different detector noise intensity $\sigma_\text{D}$ and TIC voltage resolution $u_\text{TIC}$, with a focus on changes in the short-term instability of the transmitted signal.
\begin{figure}[!h]
\includegraphics[scale = 0.85]{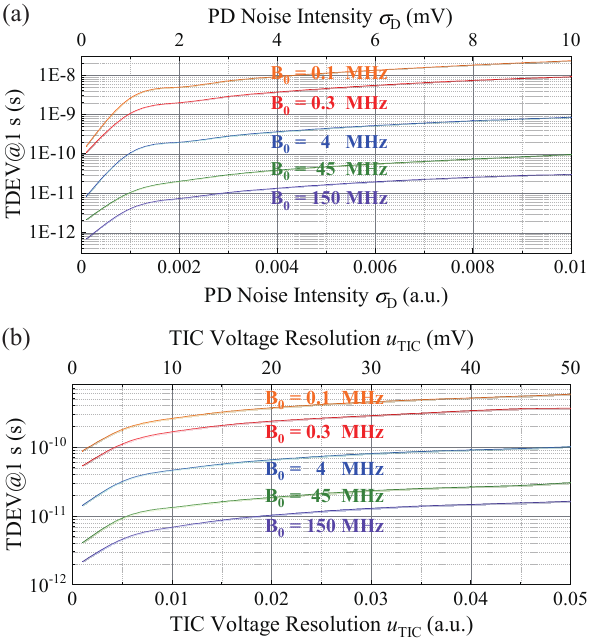}
\caption{The impact of noise on the second instability of transmitted signals under different detection bandwidths. The shaded areas on both sides of each curve show the statistical error of stability. (a) The impact of the PD noise is simulated under the parameters from $\sigma_\text{D}=0.0001$ to $\sigma_\text{D}=0.01$, corresponding to 0.1 to 10 mV. (b) The impact of the TIC voltage resolution is simulated under the parameters from $u_\text{TIC}=0.001$ to $u_\text{TIC}=0.05$, corresponding to 1 to 50 mV.}
\label{Noise_Change} 
\end{figure}

\begin{figure}[htbp]
\includegraphics[scale = 0.85]{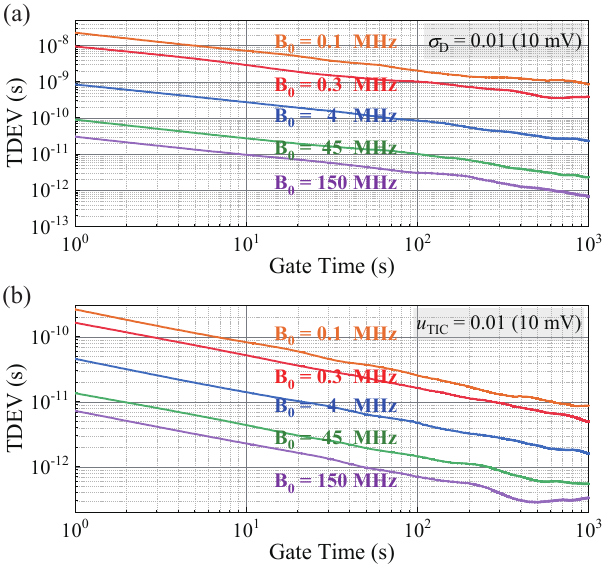}
\caption{The long-term instability of the transmitted signal with (a) PD noise intensity $\sigma_\text{D}=0.01$ and (b) voltage resolution $u_\text{TIC}=0.01$, where the number of sampling points is $N=10000$.}
\label{Noise_Change_2} 
\end{figure}

Figure~\ref{Noise_Change}(a) presents the influence of detector noise $n_\text{D}$ on the short-term instability of the transmitted signal under different detection bandwidths $B_0$. Here, the voltage resolution of TIC is set to be $u_\text{TIC}=0$. The shaded areas on both sides of each curve show the statistical error of stability under different noise intensities $\sigma_\text{D}$, with a width equal to $\pm \text{TDEV}/\sqrt{N}$\cite{804271}. Thus, we can reduce the statistical error by increasing the sampling time to obtain more sample points. It can be observed that the average short-term instability of the transmitted signal increases with the increase in noise intensity $\sigma_\text{D}$. Figure~\ref{Noise_Change}(b) illustrates the influence of TIC voltage resolution $u_\text{TIC}$ on the short-term instability of the transmitted signal under different detection bandwidths $B_0$, with the detector noise intensity set as $\sigma_\text{D}=0$. The variation trend of the short-term instability of the transmitted signal with $u_\text{TIC}$ is basically the same as that with $\sigma_\text{D}$. Moreover, we present the long-term instability of the transmitted signal under noise intensity $\sigma_\text{D}=0.01$, voltage resolution $u_\text{TIC}=0.01$ in Fig.~\ref{Noise_Change_2} respectively. The sampling point number is set to be $N=10000$, and the result shows the instabilities both decrease roughly following a $\tau^{-1/2}$ power law, close to that of the white phase noise.

\begin{figure}[htbp]
\includegraphics[scale = 0.85]{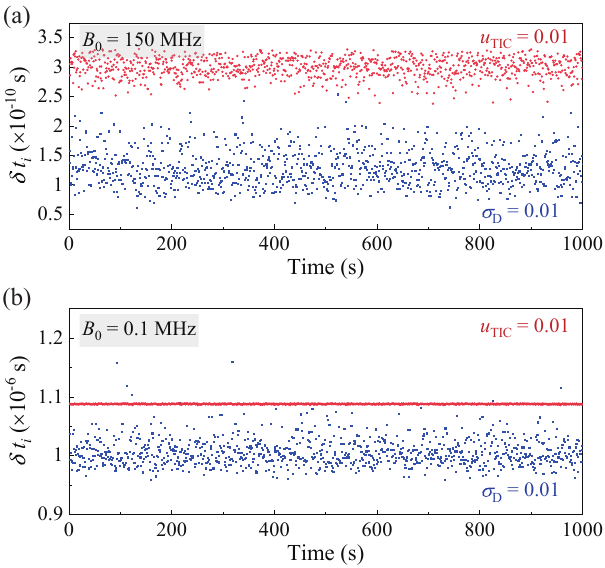}
\caption{Comparison of the distribution of time errors $\delta t_i$ caused by two types of noise after 1000 samples under the conditions of (a) $B_0=150~\text{MHz}$ and (b) $B_0=0.1~\text{MHz}$.}
\label{Noise_Change_3} 
\end{figure}

However, a comparison between the results of two types of noises shows that the TDEV in Fig.~\ref{Noise_Change}(b) is smaller under the same noise parameters, and this difference becomes more apparent as the bandwidth $B_0$ lowers. This is mainly because $u_\text{TIC}$ depicts a uniform distribution, and its parameter directly reflects the amplitude of the noise, while $\sigma_\text{D}$ is only the standard deviation of the Gaussian distribution, and there is a certain probability that the actual noise exceeds this range. Figure~\ref{Noise_Change_3}(a) specifically shows the distribution of time errors $\delta t_i$ caused by the two types of noise after 1000 samplings with $B_0=150$ MHz and $B_0=0.1$ MHz respectively. It can be seen that the peak-to-peak value of the error distribution caused by $u_\text{TIC}$ is lower than that caused by $\sigma_\text{D}$, and the smaller $B_0$ is, the greater the difference between the two. 

\subsection{Technical lower bound on the stability of the transmitted signal at different transmission distances}

In the above simulations, we have clarified the influence of different detection bandwidths $B_0$ on the detection waveform and the impact of detector noise $n_\text{D}$ and TIC voltage resolution $u_\text{TIC}$ on the final assessment of the instability of the transmitted signal under different waveforms. In this section, we will analyze the effects of the link transmission.

Equation~(\ref{Eq6}) provides the maximum effective bandwidth $B_0$ that the receiver module can offer under a certain channel length $L$. We adopt optical fiber links as the transmission channel, with a typical attenuation value of $\alpha = 0.2$ dB/km. Without loss of generality, in the following simulations, we set the amplification efficiency to $\eta = 1$, and the typical value of GBP for the photodetectors we use can reach the order of magnitude from gigahertz to 10 gigahertz. Under the experimental conditions of detector noise intensity $\sigma_\text{D} = 0.001$ and TIC voltage resolution $u_\text{TIC} = 0.01$, we sequentially simulated the cases of different detector GBP $\beta = 1$, $5$, $10$, and $50$ GHz.

\begin{figure}[b]
\includegraphics[scale = 0.4]{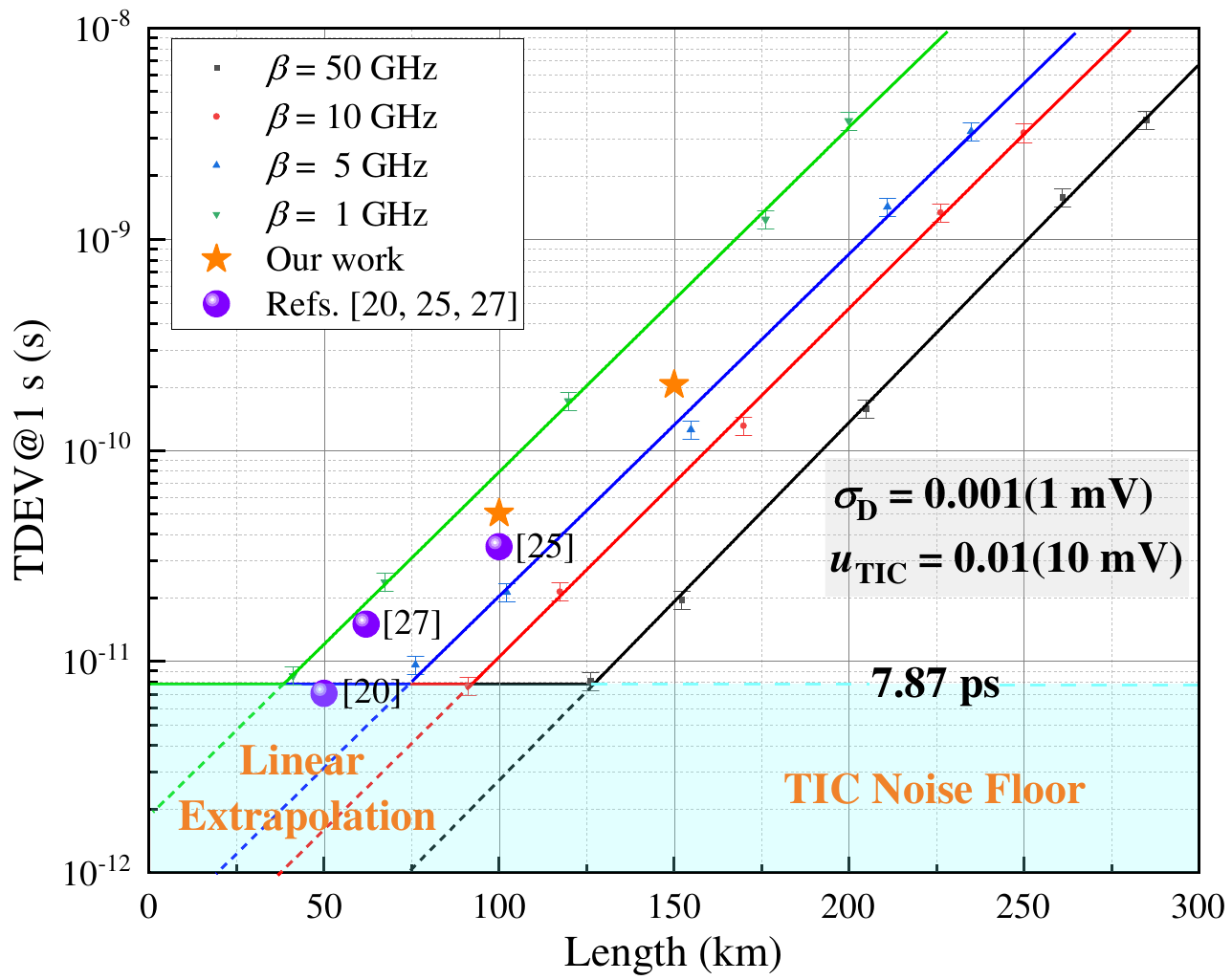}
\caption{The technical lower bound of the second stability for time transfer through a channel of length $L$ from 0 to 300 kilometers without inline amplifiers. We simulate under different values of GBP $\beta$, with the parameters of the PD noise $\sigma_\text{D}=0.001$, and the TIC voltage resolution $u_\text{TIC}=0.01$. The purple circles and yellow stars show the experimental results in both literature and our work.}
\label{Length_Change} 
\end{figure}

In Fig.~\ref{Length_Change}, the TDEV at one second exhibits a strong linear relationship with distance (correlation coefficient $r = 0.999$). The solid line in the graph represents the result of linear fitting, while the dashed portion indicates its linear extrapolation. Based on the linear fitting results, we have the empirical formula
\begin{equation}
    \text{TDEV@1 s} = c_1 \times 10^{c_2 L},
    \label{Eq26}
\end{equation}
where $c_1$ and $c_2$ are constants, and detailed derivation is provided in Appendix\ref{AppenA}. Specifically, $c_1$ represents the second instability of the signal when $L = 0$, reflecting the instability of the source-end signal measured under certain detection conditions; $c_2$ reflects the rate at which the stability of the detection signal deteriorates with distance. Some values of these parameters are listed in Table~\ref{Table1} under different GBPs.

In Fig.~\ref{Length_Change}, the results of our experiment with 100-km and 150-km non-relay transmission are marked with orange stars. The results reported in literature for time transfer without relays are indicated with purple circles\cite{RN701, RN694, RN698}, where we make an extrapolation to infer the second TDEV in Ref.~\cite{RN694}. The detector's GBP used in our experiment is close to $\beta$ = 10 GHz, while all the aforementioned points lie above the lower bound curve corresponding to $\beta$ = 5 GHz, which is consistent with our theoretical model and simulation results.
\begin{table}
\caption{\label{Table1}The values of parameters $c_1$ and $c_2$ in the empirical formula for the technical lower bound of the second instability of the transmitted signal as a function of channel length under certain values of GBP.}
\begin{ruledtabular}
\begin{tabular}{ccc}
$\beta$~(GHz) & $c_1$~(ps) & $c_2$~($\text{km}^{-1}$)\\
\hline
1 & $1.85268\pm0.14180$ & $0.01631\pm0.00024$\\
5 & $0.49047\pm0.10193$ & $0.01620\pm0.00054$\\
10 & $0.23562\pm0.02875$ & $0.01650\pm0.00029$\\
50 & $0.05448\pm0.00627$ & $0.01698\pm0.00023$\\
\end{tabular}
\end{ruledtabular}
\end{table}

Moreover, the shaded area in the figure represents the background noise of the TIC used in the experiment. This is obtained by splitting the reference electrical signal into two paths, which are then measured back-to-back by the two channels of TIC. The TDEV second stability measurement limit is approximately 7.87 ps, indicating the limiting factor for the stability measurement of the transmitted signal in the case of short distance. The dashed line within the shaded area is the linear extrapolation of the technical lower bound of the transmitted signal, representing the stability bound achievable by the short-distance transmission system if the time measurement instrument accuracy is sufficient. At the far right of the curve, when the relay-free transmission distance reaches 300 kilometers, the lower bound of stability has deteriorated to the order of 10 nanoseconds. This not only fails to meet the time transfer requirements of atomic clocks but also means that the spatial resolution capability for navigation based on time synchronization is limited to approximately 1 meter. Therefore, for high-precision time transfer over longer relay-free distances, new detection methods are needed to achieve optimization, such as adopting an energy-integrating detection scheme based on single-photon detectors.

\section{Conclusions\label{Sec4}}
In this paper, we give a theoretical model for analyzing the performance limits in single-span time transfer without inline amplifiers. We have respectively analyzed the impact of Gaussian white noise introduced by the photodetector and quantization noise introduced by the limited voltage resolution of the TIC on the instability of the measured signal under different detection bandwidths. The TDEV derived from these two noises both decrease roughly following a $\tau^{-1/2}$ power law, while the values of them differ under same values of standard deviation of Gaussian noise $\sigma_\text{D}$ and TIC resolution $u_\text{TIC}$, because of different meanings of these parameters. Moreover, by considering actual link losses, we provide an empirical formula for the relationship between the technical lower bound of the detected signal stability and the transmission distance, limited by certain GBPs. The results show that for shorter transmission distances, stability is mainly limited by the time measurement accuracy of TIC; for longer transmission distances over the 300-km scale, the precision of time synchronization is already below the 10-ns level, which is insufficient for practical applications such as navigation. Since a higher GBP results in a lower technical bound at the same transmission distance, we expect to use new detection methods with larger GBPs, such as avalanche photodetectors\cite{RN721, RN722, RN723}, to optimize long-distance high-precision time transfer under the scenarios in absence of inline amplification. Furthermore, we believe that this work offers a reference for evaluating performance limits of time transfer over different distances, and the method used in this study is general and can be adapted to provide different bounds corresponding to specific experimental schemes by adjusting the mentioned parameters. 

\begin{acknowledgments}
This work was supported by the National Natural Science Foundation of China (Grant No. 62201012) and the National Hi-Tech Research and Development (863) Program.
\end{acknowledgments}

\appendix*
\section{}
\label{AppenA}
Equation~(\ref{Eq26}) provides an empirical formula for the technical lower bound of the second instability of the transmitted signal as a function of channel length, and Table~\ref{Table1} presents the values and errors of parameters under certain conditions. In this appendix, we will outline how the derivation is carried out from the perspective of linear fitting. By linearly fitting Fig.~\ref{Length_Change}, we obtain

\begin{equation}
\log_{10}(\text{TDEV@1 s}) = (k + \Delta k) \times L + (b + \Delta b),
\label{EqA1}
\end{equation}
where $k + \Delta k$ represents the slope of the fitting line and its fitting error, and $b + \Delta b$ represents the intercept of the fitting line and its error. Further derivation yields

\begin{equation}
\text{TDEV@1 s} = 10^{(b + \Delta b)} \times 10^{((k + \Delta k) \times L)}.
\label{EqA2}
\end{equation}
Therefore, we define parameters $c_1 = 10^b$ and $c_2 = k$, leading to Eq.~(\ref{Eq26}). On the other hand, we examine the errors of the two parameters. For $c_1$, we can expand $10^{(b + \Delta b)}$ in a series with respect to the small quantity $\Delta b$, resulting in
\begin{eqnarray}
10^{(b + \Delta b)} = && 10^b + \ln(10) \times 10^b \times \Delta b \nonumber \\
&& + \frac{1}{2} \times (\ln(10))^2 \times 10^b \times (\Delta b)^2 + \cdots.
\label{EqA3}
\end{eqnarray}
Retaining terms up to the first order of expansion, we have
\begin{equation}
\Delta c_1 = 10^{(b + \Delta b)} - 10^b = \ln(10) \times 10^b \times \Delta b.
\label{EqA4}
\end{equation}

Furthermore, it's evident that $\Delta c_2 = \Delta k$. Thus, we can derive the corresponding parameters in Table~\ref{Table1} through the slope and intercept obtained from the process of linear fitting.


\providecommand{\noopsort}[1]{}\providecommand{\singleletter}[1]{#1}%
\end{document}